\documentclass[10pt]{iopart}
\usepackage[utf8]{inputenc}
\usepackage{graphicx}
\usepackage{epstopdf}
\usepackage[normalem]{ulem}

\usepackage{iopams}

\begin{document}

\title[Flowing skyrmions]{Dynamics of flowing 2D skyrmions}

\author{Rodrigo C. V. Coelho, Mykola Tasinkevych, Margarida M. Telo da Gama }

\address{Centro de F\'{i}sica Te\'{o}rica e Computacional,  Faculdade de Ci\^{e}ncias, Universidade de Lisboa, 1749-016 Lisboa, Portugal\\ Departamento de F\'{\i}sica, Faculdade de Ci\^{e}ncias, Universidade de Lisboa, P-1749-016 Lisboa, Portugal.
}

\ead{rcvcoelho@fc.ul.pt}
\vspace{10pt}

\begin{abstract}
We investigate, numerically, the effects of externally imposed material flows on the structure and temporal evolution of liquid crystal skyrmions. 
The dynamics of a 2D system of skyrmions is modeled using the Ericksen-Leslie theory, which is based on two coupled equations, one for material flow and the other for the director field. As the time scales of the velocity and director fields differ by several orders of magnitude for realistic values of the system parameters, we have simplified the calculations by assuming that the velocity relaxes instantaneously when compared to the relaxation of the director field. Thus, we have used a finite-differences method known as artificial compressibility with adaptive time step to solve the velocity field and a fourth-order Runge-Kutta method for the director field. 
We characterized the skyrmion shape or configuration as a function of the time and the average velocity of the flow field. We found that for velocities above a certain threshold, the skyrmions stretch in the direction perpendicular to the flow, by contrast to the regime of weak flows where the skyrmions stretch along the streamlines of the flow field. These two regimes are separated by an abrupt (first-order) dynamical transition, which is robust with respect to e.g., the liquid crystal elastic anisotropy. Additionally, we have found how the presence of a second skyrmion affects the evolution of the shape of the skyrmions, by comparing the evolution of pairs of skyrmions to the evolution of a single-skyrmion. 
\end{abstract}

%
%
%
%
\ioptwocol
%


\section{Introduction}

Active colloids represent a new class of nonequilibrium soft matter in which energy harvesting and conversion take place at the level of the individual particles. In the last decade, significant progress has been made in developing synthetic self-propelled micro-particles, which are capable of converting the free energy of the environment into mechanical energy of translational or rotational motion~\cite{Bechinger2016}.  This offers the possibility to use such particles as autonomous micromotors \cite{Baraban2012,Soto:2021}  for, e.g. delivering drugs \cite{Patra2013,Xu2020}, sensing specific substances \cite{Wu:2010}, assembling structures via the autonomous local deposition of materials \cite{Sanchez2015}, or removal of contaminants from water \cite{Soler2013,Soler2014}. From a basic science point of view, active colloids exhibit novel types of emergent collective behavior, not observed in passive colloids, such as ``living crystals''  that are mobile, break apart and reform again \cite{Palacci2013} or motility-induced phase separation in systems with purely repulsive interactions \cite{Fily2012, Redner2013, Levis2014}.

Recently, a novel class of soft active matter has been realized experimentally, where topological solitons in confined chiral liquid crystals (LCs)  are the elementary building blocks of the active matter system \cite{Ackerman2017}. These solitons, named ``skyrmions'' are spatially localized, non-singular configurations of the LC director field that cannot be transformed continuously into the uniform state. They are low-dimensional analogs of Skyrme solitons in nuclear physics \cite{Skyrme1962}. Both three-dimensional (3D) \cite{Smalyukh2010} and two-dimensional (2D) \cite{Ackerman2014} skyrmions have been realized. The core of a 3D skyrmion is a double twist torus, where the director twists from the torus axis in all orthogonal (to the torus axis) directions \cite{Smalyukh2010}.  Experiments and numerical calculations based on the Frank-Oseen elastic free energy \cite{Tai2018} reveal a rich structural behavior and conformational transitions between skyrmion states with the same or different Hopf indices. By contrast to active colloids which are solid, LC skyrmions are soft as they lack physical interfaces and their motion is accompanied by the periodic expansion and contraction of topology-protected distorted LC regions, mimicking the behavior of biological cells.

The motion of the LC skyrmions is powered by a time-dependent electric field applied to the LC in a direction normal to the confining surfaces -- a set up that resembles the one used in LC display technology  \cite{Ackerman2017}. The basic physical mechanism of the skyrmion motion is related to the ``non-reciprocal'' rotational dynamics of the LC director field when the electric field is turned on and off. Surprisingly, it is possible to control both the speed and the direction of the motion by varying the strength and the modulation frequency of the applied electric field \cite{Ackerman2017}. Additionally, skyrmion motion can be controlled by taking advantage of the unique optical properties of LCs. For example, the size and velocity of solitons, as well as their collective dynamics and self-assembly can be controlled by combining laser tweezer techniques and photo-patterning of the in-plane LC director \cite{Sohn2020}.

LC skyrmions exhibit effective elastic interactions that can be easily tuned in strength or switched from attractive to repulsive \cite{Sohn2019}. When the voltage modulation period is shorter than the LC response time, the skyrmion interactions are intrinsically out-of-equilibrium, resulting in remarkably rich emergent collective dynamics with reconfigurable out-of-equilibrium assemblies of skyrmions. At high packing fractions, hexagonal crystallites of tightly packed solitons can be brought into coherent motion along an arbitrary direction, which leads to an increased hexatic order parameter and is accompanied by the anisotropic deformation of the hexagonal soliton lattice \cite{Sohn2020}. Active skyrmions can also be used to entrap \cite{Porenta2014} and transport  microparticles \cite{Sohn2018}, which provides opportunities for the development of novel electro-optic responsive materials as the experimental conditions for active skyrmions are similar to those used in LC display technologies. 

Despite the extensive body of experimental research, the many-body dynamics of LC solitons remains poorly understood. Existing numerical investigations are limited to a very small number of skyrmions and exploit the relaxation dynamics of the LC director field only in order to understand the field-induced motion of the skyrmions \cite{Ackerman2017,Sohn2019}, ignoring completely the effects of the material flow field.  Experiments revealed the presence of weak backflows associated with the skyrmion motion \cite{Ackerman2017}, but no systematic study of this effect was pursued. On the other hand, detailed numerical analysis of the dynamics of nematic LCs subject to step-like voltage modulations \cite{Kos2020} demonstrated robust generation of material flows by dynamic electric fields using the backflow effects. Additionally, in future lab-on-a-chip applications, active particles will undoubtedly encounter shear flows and will need to autonomously sense and respond to them.  

In the present study, we focus on the effects of externally imposed material flows on the structure and temporal evolution of LC skyrmions. In particular, we obtain the skyrmion shape or configuration as a function of time and the average velocity of the flow field. Surprisingly, at early times and for velocities above a certain threshold, the skyrmions stretch in the direction perpendicular to the flow, by contrast to the regime of weak flows where the skyrmions are stretched along the streamlines of the flow. These regimes are separated by an abrupt (first-order) dynamical transition, which is robust with respect to e.g., changes of the LC elastic constants. Additionally, we show that the presence of a neighboring skyrmion significantly affects the evolution of the skyrmion shape when compared to the single-skyrmion case. In Sec.~\ref{method-sec} we outline the theory used to describe the dynamics of the system in 2D and the numerical method used to solve the dynamical equations, for the set of material parameters under consideration. In Sec.~\ref{results-sec} we present and discuss the results for single skyrmions under flow and and for pairs of skyrmions. We emphasise the results for the configuration transition found for single skyrmions under flow. Finally, in Sec.~\ref{conclusion-sec} we conclude and point directions for future work. 

\section{Theory and numerical method}
\label{method-sec}

In this section we describe the equations used to model the skyrmion dynamics and the numerical methods employed to solve them. We consider a chiral nematic LC under confinement, far from any bulk transition, reducing the skyrmion ordering dynamics to the dynamics of the director field (i.e., we assume that the scalar nematic order parameter is constant throughout the sample and 
does not contribute to the ordering dynamics).  

\subsection{Ericksen-Leslie dynamics} 

Liquid crystals are materials that flow like liquids, but are composed of non-spherical particles with a preferential direction of alignment in the nematic phase, known as the director. Cholesterics are twisted nematics, where the director rotates over a characteristic distance, known as the pitch. The simplest model to describe the dynamics of the director field, which is adequate deep in the nematic or the cholesteric phase, was proposed by Ericksen and Leslie~\cite{Ericksen1962, doi:10.1098/rspa.1968.0195}. It consists of two equations: one for the material flow and the other for the director field.

For the material flow, we use the Navier Stokes equation together with the continuity equation:
\begin{eqnarray}
 &&\rho \partial_t u_\alpha + \rho u_\beta \partial_\beta u_\alpha = \partial_\beta \left[ - P \delta_{\alpha\beta} + \sigma_{\alpha\beta}^{v}   + \sigma_{\alpha\beta}^{e}   \right] \label{NS-eq}\\
 && \partial_\alpha u_\alpha = 0\label{cont-eq},
\end{eqnarray}
where the viscous stress tensor is:
\begin{eqnarray}
 \sigma_{\alpha\beta}^{v} = && \alpha_1 n_\alpha n_\beta n_\mu n_\rho D_{\mu\rho}+ \alpha_2 n_\beta N_\alpha + \alpha_3 n_\alpha N_\beta \nonumber \\
&&+ \alpha_4 D_{\alpha\beta} + \alpha_5 n_\beta n_\mu D_{\mu\alpha}  + \alpha_6 n_\alpha n_\mu D_{\mu\beta} .
\end{eqnarray}
Here $\rho$ stands for the fluid density, $P$ for the hydrostatic pressure, $\mathbf{u}$ for the fluid velocity, $\mathbf{n}$ for the director field (unit vector in the direction of preferential alignment of the molecules) and $\alpha_n$'s for the Leslie viscosities of the material. The kinematic transport, which represents the effect of the macroscopic flow field on the microscopic structure, is given by:
\begin{eqnarray}
 N_\beta = \partial_t n_\beta + u_\gamma \partial_\gamma n_\beta - W_{\beta \gamma} n_\gamma
\end{eqnarray}
while the strain and vorticity tensors are, respectively:
\begin{eqnarray}
 D_{\alpha\mu} = \frac{1}{2}\left (  \partial_\alpha u_\mu + \partial_\mu u_\alpha \right), \: W_{\alpha\mu} = \frac{1}{2}\left ( \partial_\alpha u_\mu - \partial_\mu u_\alpha \right).
\end{eqnarray}
The elastic stress tensor is:
\begin{eqnarray}
 \sigma_{\alpha\beta}^{e} = -\partial_\alpha n_\gamma \frac{\delta \mathcal{F}}{\delta (\partial_\beta n_\gamma)},
\end{eqnarray}
where $\mathcal{F}$ is the Frank-Oseen elastic free energy:
\begin{eqnarray}
 \mathcal{F} =&& \int dV \Big\{        \frac{K_{11}}{2} (\nabla \cdot \mathbf{n})^2 + \frac{K_{22}}{2} [\mathbf{n}\cdot (\nabla \times \mathbf{n}) + q_0 ]^2 \\ && +  \frac{K_{33}}{2}[  \mathbf{n}\times (\nabla \times\mathbf{n}) ]^2       \Big\} .
 \label{free-energy-eq}
\end{eqnarray}
$K_{11}$, $K_{22}$, $K_{33}$ are the LC elastic constants, and $q_0=2\pi/p$, with $p$ the cholesteric pitch. 
The second equation describes the time evolution of the director field:
\begin{eqnarray}
 \partial_t n_\mu = \frac{1}{\gamma} h_\mu - \lambda n_\alpha D_{\alpha\mu} - u_\gamma \partial_\gamma n_\mu + W_{\mu\gamma}n_\gamma,
 \label{director-time-eq}
\end{eqnarray}
where $\gamma=\alpha_3-\alpha_2$ is the rotational viscosity, $\lambda=(\alpha_3+\alpha_2)/(\alpha_3-\alpha_2)$ is the aligning parameter, with $\vert \lambda\vert>1$ for flow aligning particles and $\vert \lambda\vert<1$ for flow tumbling ones. Finally, the molecular field is: 
\begin{eqnarray}
 h_\mu = -\frac{\delta \mathcal{F}}{\delta n_\mu}.
\end{eqnarray}

\subsection{Parallel plates modeling}
\label{plates-sec}

We consider 2D domains and assume that the system is invariant in the direction perpendicular to the plane. This is a simplification used  to reduce the computational cost, which is quite high for 3D simulations using the numerical techniques that will be discussed in the next section. To this end, we add to the free energy density of Eq.~(\ref{free-energy-eq}) an effective anchoring term everywhere, which mimics the anchoring of the parallel plates as was done, for instance, in Ref.~\cite{C9SM02312G}:
\begin{eqnarray}
 f_W = - \frac{W_0}{2} (\mathbf{n_w}\cdot\mathbf{n})^2 ,
\end{eqnarray}
where $W_0$ is the anchoring strength and $n_w$ is the normal to the plane. In 3D, the anchoring is applied only at the plates, see Ref.~\cite{PhysRevE.101.042702}.

In addition, we add a friction force to the right hand side of Eq.~(\ref{NS-eq}) to describe the resistance to the fluid flow caused by the parallel plates:
\begin{eqnarray}
 \mathbf{F} = -\chi \mathbf{u},
\end{eqnarray}
where $\chi$ is the friction coefficient. Assuming Poiseuille flow, $u(z) = -a/(2\nu) z(z-L)$, due to an external acceleration $\mathbf{a}$, the average velocity in the direction of the acceleration is $ \langle u \rangle = \frac{a L^2}{12 \nu}$. Thus the 2D friction coefficient corresponding to the same average velocity in a 3D system of parallel plates separated by $L$ is 
\begin{eqnarray}
 \chi = \frac{12\mu}{L^2},
\end{eqnarray}
where $\mu=\rho \nu$ is the absolute viscosity of the fluid and $\nu$ is the kinematic viscosity. 
In simulation units (see Table~\ref{tab1}), we consider $L=34\Delta x$, which is close to the cholesteric pitch, giving $\chi = 4.25$.

\subsection{Material parameters and time scales}

The elastic constants of a typical LC such as MBBA at 22$^\circ$C~\cite{de1993physics} are $K_{11}=5\times 10^{-12}$ N, $K_{22}/K_{11}=0.42$ and $K_{33}/K_{11}=1.4$. However, in the simulations that follow we have used smaller elastic constants (52.4 times smaller, keeping the ratios) in order to enhance numerical stability. The parameters in simulation units are given in the Appendix and the following analysis of the characteristic times uses the actual parameters. 

The Leslie viscosities are $\alpha_4=0.08$ Pa.s, $\alpha_1 = 0.08 \alpha_4$, $\alpha_2 = -0.93\alpha_4$, $\alpha_3 = -0.014 \alpha_4$, $\alpha_5 = 0.56 \alpha_4 $ and $\alpha_6 = -0.41 \alpha_4$.

Following Ref.~\cite{Svenek2001}, the characteristic time scale for the relaxation of the director field is
\begin{eqnarray}
 \tau_n = \frac{\gamma L^2}{K}
\end{eqnarray}
where $L$ is the relevant length scale. In our system, it is the separation between the parallel plates but it could be the electric correlation length in systems under an external electric field. On the other hand, the characteristic time scale for fluid flow reads:
\begin{eqnarray}
 \tau_v = \frac{\rho L^2}{\mu},
\end{eqnarray}
with $\mu=\alpha_4/2$. The ratio between these time scales measures the unsteadiness of the flow which is $\frac{\tau_v}{\tau_n} \sim 10^{-6}$ for the parameters given above. In problems involving strong electric fields, the time scales may become comparable~\cite{PhysRevE.89.032508}. In the present setting, however, the time scales are so different that we can consider that the fluid relaxes instantaneously when compared to the director field relaxation. Therefore we can obtain the fluid velocity steady state solution while keeping the director field fixed. Additionally, the Reynolds number is very small, $Re\sim 10^{-6}$ and thus we can set the left hand side of Eq.~(\ref{NS-eq}) to zero. 

One relevant non-dimensional number that characterize the flow is the Ericksen number:
\begin{eqnarray}
 Er = \frac{U L \mu}{K},
\end{eqnarray}
which gives the ratio of the viscous over the elastic forces.
In the simulations that follow, we will vary this number by changing the average velocity of the fluid, due to an external acceleration $\mathbf{a}$. The analysis of Sec.~\ref{plates-sec}, shows that the resulting velocity is $U = a/\chi$. A second non-dimensional number also relevant in the problem is the ratio of the anchoring over the elastic forces~\cite{PhysRevE.101.042702}:
\begin{eqnarray}
 N_W = \frac{W_0 p^2}{K}.
\end{eqnarray}
In most simulations this anchoring number was kept fixed (except in simulations of Fig.~\ref{ericksen-fig}).

\begin{figure*}[h]
\center
\includegraphics[width=\linewidth]{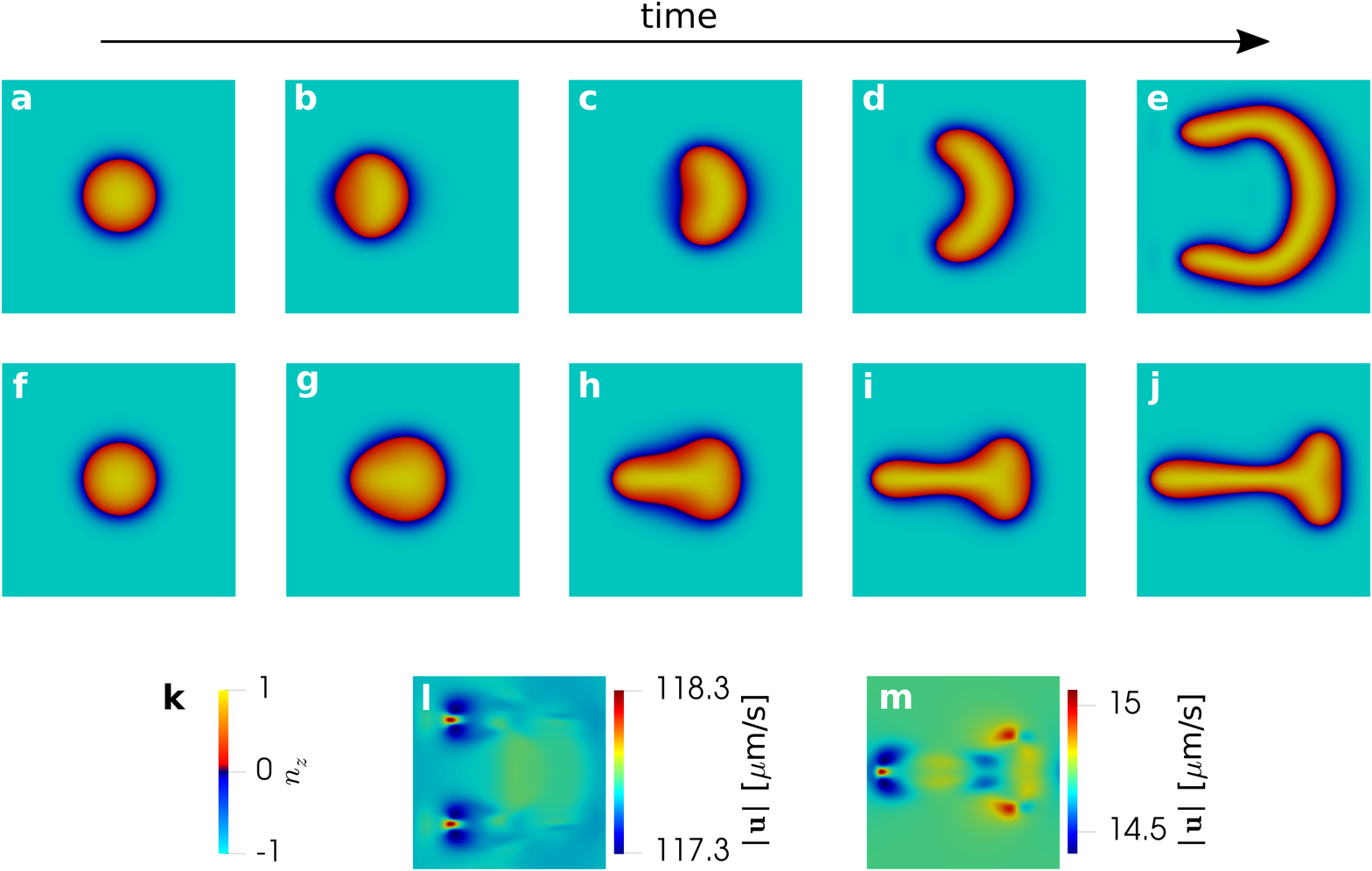}
\caption{ Time evolution of the skyrmions for two different average velocities. On the top row, the average velocity is $\langle u \rangle=117.64\, \mu$m/s, and the time of the different frames is: (a) 0 s, (b) 9.5 s, (c) 21.5 s, (d) 46 s, (e) 90 s. In the middle row, the average velocity is $\langle u \rangle=14.71\, \mu$m/s, and the time of the different frames is: (f) 0 s, (g) 24.5 s, (h) 54.5 s, (i) 79 s, (j) 106.5 s. (k) is the color bar for the frames (a) to (j). (l) and (m) depict the magnitude of the velocity field corresponding to the frames (e) and (j) respectively. }
\label{examples1-fig}
\end{figure*}

\subsection{Numerical implementation}

Using the approximations discussed in the previous section, Eq.~(\ref{NS-eq}) becomes
\begin{eqnarray}
 \partial_\beta \left[ - P \delta_{\alpha\beta} + \sigma_{\alpha\beta}^{v}   + \sigma_{\alpha\beta}^{e}   \right] - \chi u_\alpha  =0.
 \label{NS-mod-eq}
\end{eqnarray}
We solve Eqs.~(\ref{NS-mod-eq}) and~(\ref{cont-eq}) using the artificial compressibility method~\cite{CHORIN196712}. This is a finite-differences method to obtain steady state solutions of the Navier-Stokes equation. It considers a pseudo time $\tau$, and the solution is iterated until the flow reaches the steady state. In the case under study, the equations are:
\begin{eqnarray}
 &&\partial_\tau u_\alpha = \partial_\beta \left[ - P \delta_{\alpha\beta} + \sigma_{\alpha\beta}^{v} + \sigma_{\alpha\beta}^{e}   \right] - \chi u_\alpha    \label{pseudo-time-eq1}\\
 && \partial _\tau P = - c^2 \partial_\alpha u_\alpha \label{pseudo-time-eq2},
\end{eqnarray}
where $c$ is an arbitrary constant. The previous equations are iterated until $\partial _\tau P \rightarrow 0$ which ensures that the continuity equation is satisfied. It is also possible to control the convergence through the velocity field rather than the pressure, which was done in this work. We consider that convergence is obtained when
\begin{eqnarray}
 \textrm{max}\left(\frac{ \left\vert \vert\mathbf{u}_{\textrm{\tiny{new}}}\vert  -  \vert\mathbf{u}_{\textrm{\tiny{old}}} \vert   \right\vert }{ \langle   \vert\mathbf{u}\vert \rangle }\right) < 10^{-6},
\end{eqnarray}
which guarantees that the fluid velocity has converged locally.  Here, $\mathbf{u}_{\textrm{\tiny{new}}}$ and $\mathbf{u}_{\textrm{\tiny{old}}}$ are the velocity fields in the current and the previous pseudo time steps respectively, while $\langle   \vert\mathbf{u}\vert \rangle$ is the average of the magnitude of the velocity in the entire domain. The pseudo time step $\Delta\tau$ is chosen using an adaptive scheme in order to speed up convergence~\cite{10.5555/148286}. In short, we calculate the velocity field twice: first by using a time step $\Delta\tau$, which gives $\mathbf{u}_{\Delta\tau}$, and then by using half of this time step  $\Delta\tau/2$ twice, which gives $\mathbf{u}_{\Delta\tau/2}$. We compute the difference of the local magnitude of the velocity calculated using these two time steps and choose the next time step as follows: 
\begin{eqnarray*}
 \Delta\tau_{\textrm{\tiny{new}}} = 0.9 \Delta\tau_{\textrm{\tiny{old}}} \textrm{min} \left( \textrm{max}\left( \left(  \frac{\textrm{tol}}{2 \Delta u_{\textrm{\tiny{max}}}} \right)^{\frac{1}{2}}    ,0.3 \right) , 2  \right),
\end{eqnarray*}
where we set tol $=10^{-6}$ and 
\begin{eqnarray}
 \Delta u_{\textrm{\tiny{max}}} = \textrm{max}\left( \frac{\left\vert \vert\mathbf{u}_{\Delta\tau}\vert - \vert\mathbf{u}_{\Delta\tau/2}\vert \right\vert}{\langle \vert\mathbf{u}\vert  \rangle} \right).
\end{eqnarray}
The 0.9 is a safety factor to increase the chances of success in the next iteration, i.e. $\Delta u_{\textrm{\tiny{max}}} <$ tol, and the maximum and minimum values are used to prevent extreme changes in successive time steps.  

The spatial derivatives on the r.h.s of Eq.~(\ref{director-time-eq}) are approximated by using finite-differences and the integration over time is performed using the fourth-order Runge-Kutta method. More specifically, at each time step in  Eq.~(\ref{director-time-eq}) we use the steady state (with respect to the pseudo time $\tau$) fluid velocity field obtained by solving  Eqs.~(\ref{pseudo-time-eq1}) and~(\ref{pseudo-time-eq2}), with the director field kept constant. Typically, the steady state fluid velocity is achieved after a few hundred pseudo time steps $\Delta \tau$ (per time step of the director field evolution). At each time step, the initial guess of the fluid velocity field, is the velocity field of the previous time step, which increases the speed of convergence. Although this scheme is computationally much faster than the co-evolution of the fluid and the director fields with the same time step, it is still very costly for 3D simulations. Typically, in our 2D simulations, it takes 0.08 MLUPS (million lattice updates per second) in 8 threads. In addition, the simulation of 3D skyrmion, or toron~\cite{PhysRevE.101.042702}, faces another challenge: strong spurious currents appear close to the two point defects which ``decorate'' the skyrmions at their lower and upper regions close to the confining surfaces. This is due to the poor resolution of the defect cores with the current method based on a uniform rectangular grid. More advanced adaptive mesh techniques are required in order to resolve the regions around these point defects, which in turn call for more sophisticated numerical techniques to solve the fluid flow field in 3D problems with realistic parameters, in reasonable time.

\section{Results}
\label{results-sec}

\subsection{One skyrmion}
\label{1sk-sec}

We start by describing the dynamics of a single skyrmion on a square domain of size $40 \mu$m and periodic boundary conditions. The initial configuration is set up as follows. We begin with an Ansatz for the director field as in Ref.~\cite{Foster2019}: 
\begin{eqnarray}
 &&n_y = \sin(a) \cos(mb + g)\nonumber\\
 &&n_x = \sin(a) \sin(mb + g)\nonumber\\
 &&n_z = -\cos(a),
\end{eqnarray}
where
\begin{eqnarray}
 &&a =  \frac{\pi}{2}\left[1 - \tanh\left(\frac{B}{2}(r-R)\right)\right]\\
 && b = \tan^{-1}\left(\frac{x-C_x}{y-C_y}\right)\\
 && r=\sqrt{(x-C_x)^2+(y-C_y)^2} .
\end{eqnarray}
The parameter $R$ controls the size of the skyrmion, $B$ controls the sharpness of the interface that separates the inner and outer regions, $m$ is the winding number of the skyrmion, $g$ controls the direction of the skyrmion, $r$ is the distance from the skyrmion center and $b$ is the 2D polar angle. The values of the parameters used in the simulation are (in simulation units): $m=1$, $g = \pi/2$, $R=0.7p$, $B=0.5$, $C_x=L_X/2$, $C_y=L_Y/2$. The velocity field is set to zero while the the Ansatz configuration is relaxed until it reaches the steady state.

Then a body force is applied to set up a mass flow (from left to right in the figures). We observe that the velocity of the skyrmion's center of mass is approximately the same as the average velocity in the domain, meaning that the skyrmion is simply advected by the fluid flow. Small differences between these velocities are observed due to the change in the skyrmion's center of mass as a result from its shape change, as discussed next. 

The results of the simulations suggest that the configuration of the skyrmion changes with time and does not reach a steady state. The skyrmion keeps stretching with time until it approaches the domain size and then interacts with its mirror image(s) due to the periodic boundaries. More interestingly, the skyrmion's shape is significantly different below and above a threshold velocity $\langle u \rangle_{th}$. In Fig.~\ref{examples1-fig}, we illustrate the time evolution of the skyrmion configuration for two different velocities. At large velocities, for $\langle u \rangle>\langle u \rangle_{th}$, the skyrmion stretches in the direction perpendicular to the flow and acquires a ``C'' shape. When it reaches the border of the domain, its curvature increases due to the periodic boundaries. By contrast, at low velocities, for $\langle u \rangle<\langle u \rangle_{th}$, the skyrmion stretches in the direction of the flow, acquiring a ``T'' shape. Figs.~\ref{examples1-fig} (l) and (m) illustrate the velocity field in the last step for each velocity (Figs.~\ref{examples1-fig} (e) and (j)). Notice that the velocity is essentially uniform except close to the edges of the skyrmions where it is slightly different.

\begin{figure}[h]
\center
\includegraphics[width=\linewidth]{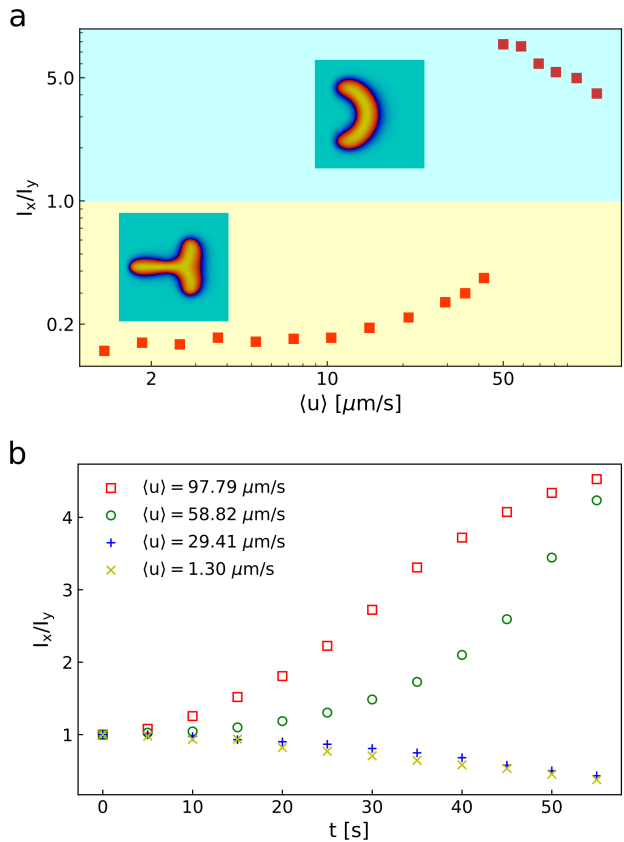}
\caption{(a) Configuration transition characterized by the quantity $I_x/I_y$ as a function of the average velocity. We observe a discontinuous transition at $\langle u \rangle_{th}\approx45.96\, \mu$m/s above which the skyrmion is stretched in the direction perpendicular to the flow. The measurements are taken when the length or the height of the skyrmion reaches 25 $\mu$m. (b) Time evolution of the quantity $I_x/I_y$ for different velocities.}
\label{ix-iy2-fig}
\end{figure}

\begin{figure}[h]
\center
\includegraphics[width=\linewidth]{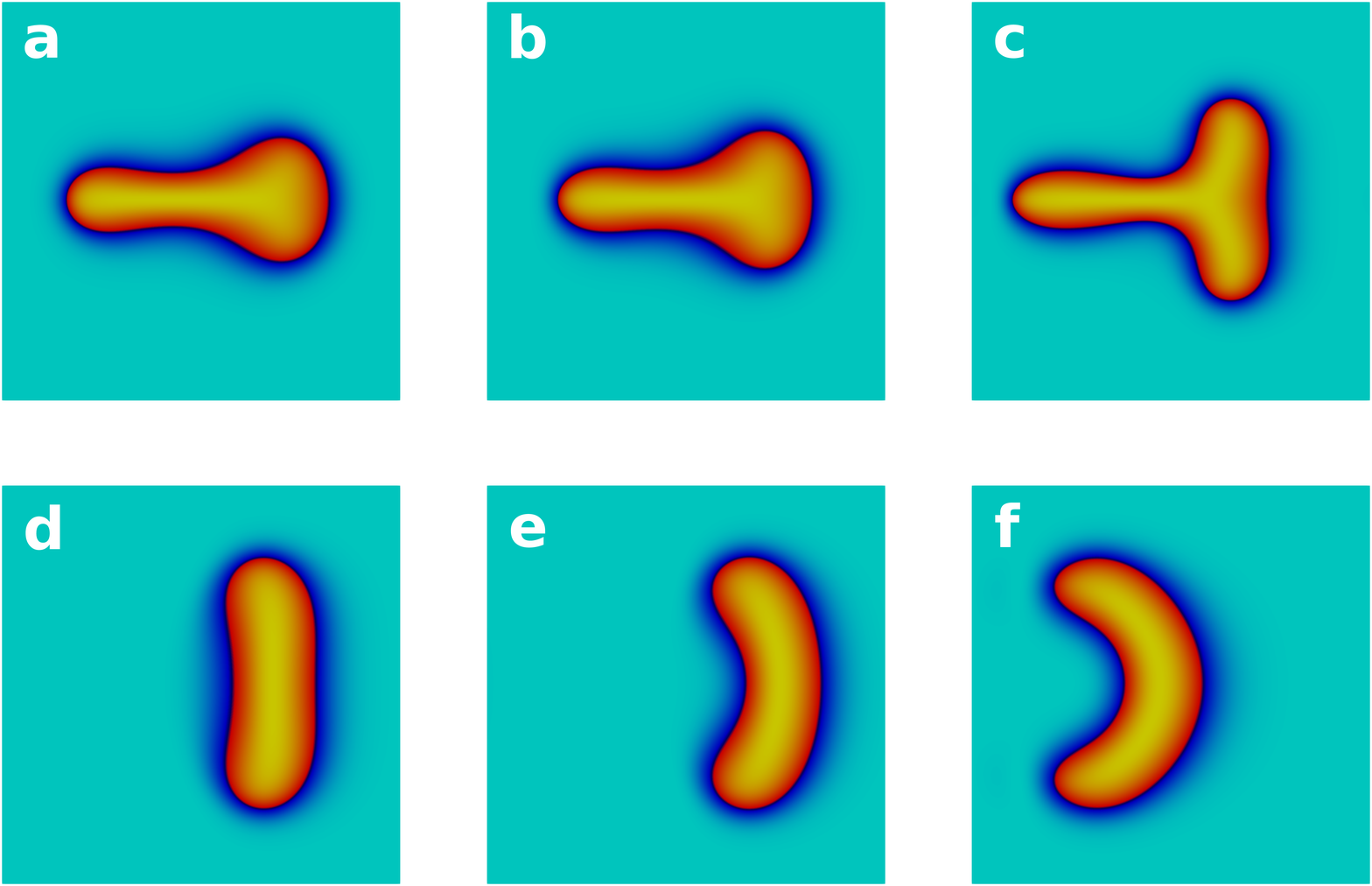}
\caption{ Comparison of skyrmions with length or height equal to 25 $\mu$m at different average velocities. On the top, skyrmions stretched in the direction of the flow with average velocity (a) $\langle u \rangle=1.30\, \mu$m/s, (b) $\langle u \rangle=10.4\, \mu$m/s, (c) $\langle u \rangle=14.71\, \mu$m/s. On the bottom, skyrmions stretched in the direction perpendicular to the flow with average velocity (d) $\langle u \rangle=50.0\, \mu$m/s, (e) $\langle u \rangle=58.82\, \mu$m/s, (f) $\langle u \rangle=117.65\, \mu$m/s.    }
\label{examples2-fig}
\end{figure}

In order to analyse the configuration transition, we define an order parameter as the second moment of the skyrmion spatial distribution (akin to mass in the moment of inertia):
\begin{eqnarray}
&&I_x = \int H(n_z) (y-y_{CM})^2 dA \\
&&I_y = \int H(n_z) (x-x_{CM})^2 dA,
\label{ix-it-eq}
\end{eqnarray}
where $H(x)$ is the Heaviside step function, $x_{CM}$ and $y_{CM}$ are the coordinates of the ``center of mass'' of the skyrmion defined as $x_{CM} = \int H(n_z)\, x\, dA/(L_X\, L_Y)$ and $y_{CM} = \int H(n_z) \,y \,dA/(L_X\, L_Y)$. The integration is over the entire area of the domain. The ratio $I_x/I_y$ quantifies the skyrmion distribution along the directions perpendicular and parallel to the flow. If $I_x/I_y>1$ the skyrmion is stretched perpendicular to the flow while if $I_x/I_y<1$ the skyrmion is stretched parallel to it. Fig.~\ref{ix-iy2-fig}(a) shows this order parameter as a function of the fluid velocity. As the configuration changes with time, (see Fig.~\ref{ix-iy2-fig}(b)) we measured the quantity $I_x/I_y$ when the length or the height of the skyrmion reaches $80\Delta x = 25\, \mu$m. The results reveal that the configuration transition occurs at a velocity $\langle u \rangle_{th}\approx45.96\, \mu$m/s, corresponding to an Ericksen number $Er \approx 192.7$, where we used the pitch as the characteristic length. We illustrate in Fig.~\ref{examples2-fig} the configuration of the skyrmions at the time when its length or height reaches $80\Delta x = 25 \mu$m. For velocities below the threshold, the length of the ``T'' cross increases as the velocity increases. For velocities above the threshold, the skyrmion becomes more curved as the velocity increases with the ``C'' closing up. A full analysis of the forces acting on the skyrmions is difficult as the director field and the shear stresses are far from uniform. Additional complications result from the effect of the substrate friction and the anchoring by the plates although these forces are likely to be sub-dominant. 

\begin{figure}[h]
\center
\includegraphics[width=\linewidth]{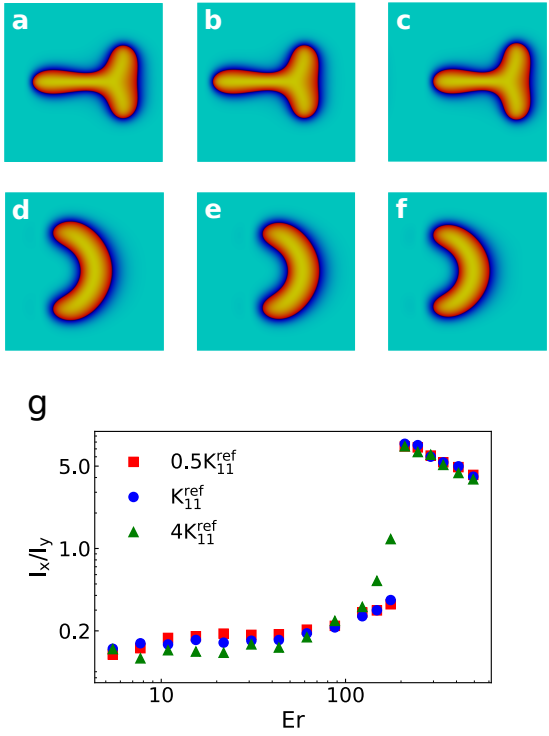}
\caption{Shape similarity of skyrmions with the same Ericksen and anchoring numbers for different elastic constants and velocities. Figures (a), (b) and (c) correspond to $Er=123.3$ and (d), (e) and (f) to $Er=493.4$. The times were chosen by inspection for skyrmions with a similar size. (a) $K_{11}= 0.5 K_{11}^{ref}$ and $\langle u\rangle=14.7 \mu$m/s.  (b) $K_{11}= K_{11}^{ref} = 9.54\times 10^{-14}$ N and $\langle u\rangle=29.4 \mu$m/s. (c) $K_{11}=4K_{11}^{ref}$  and $\langle u\rangle=117.6 \mu$m/s. (d) $K_{11}= 0.5 K_{11}^{ref}$ and $\langle u\rangle=58.8 \mu$m/s. (e) $K_{11}= K_{11}^{ref}$ and $\langle u\rangle=117.6 \mu$m/s. (f) $K_{11}= 4 K_{11}^{ref}$ and $\langle u\rangle=470.4 \mu$m/s. (g) Relation between the quantity $I_x/I_y$ and the Ericksen number for the three different elastic constants. }
\label{ericksen-fig}
\end{figure}

We have verified that skyrmions with different elastic constants and flow velocities but with the same Ericksen number (ratio of viscous to elastic forces) have identical shape. 
In Fig.~\ref{ericksen-fig}, we depict skyrmions for velocities below and above the threshold and different elastic constants: one half of those used earlier (Fig.~\ref{ericksen-fig}(a) and (d)), those used earlier (or the reference elastic constants, represented as $K_{11}^{ref}$ in the figures) (Fig.~\ref{ericksen-fig}(b) and (e)) and four times larger than the reference elastic constants (Fig.~\ref{ericksen-fig}(c) and (f)). Note that we have also changed the anchoring strength to keep the anchoring number $N_W=12.3$ constant. As the shape of the skyrmions changes with time, we chose an instant when the size of the skyrmions is similar. We find that the shape of the skyrmions for the same Ericksen ($Er$) and anchoring ($N_W$) numbers is very similar.
We have checked the collapse of the configuration diagrams (similar to the one shown in Fig.~\ref{ix-iy2-fig}(a)) with the Ericksen number, obtained for different values of the elastic constants. As shown in Fig.~~\ref{ericksen-fig}(g), the configuration transition occurs for the same value of $Er$ in the three cases.
 
\begin{figure}[h]
\center
\includegraphics[width=\linewidth]{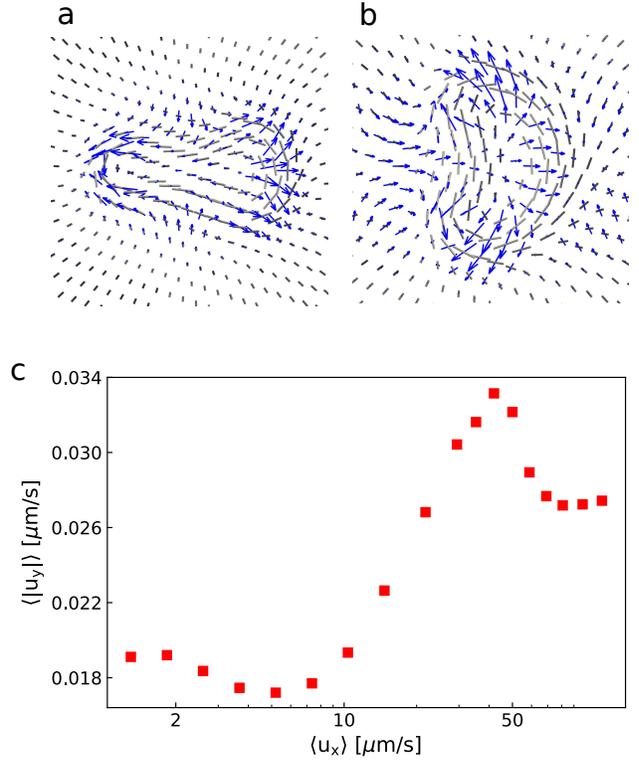}
\caption{Anisotropy of the flow field. (a) and (b) show two examples of the flow field close to the skyrmion.
The average velocities are (a) $\langle u \rangle =1.3 \mu$m/s and (b) $\langle u \rangle =117.6\mu$m/s.
The gray lines represent the director field while the blue arrows represent the fluid velocity relative to the skyrmion motion. (c) Norm of the $y$-component against the $x$-component of the velocity field, both averaged in space. The measurements are taken when the length or the height of the skyrmion reaches 25 $\mu$m. }
\label{flow-y-fig}
\end{figure}

We have also checked that the skyrmion configuration transition persists when the elastic anisotropy~\cite{coelho2020director} of the LC is switched off (single LC elastic constant) and thus conclude that the mechanism is dominated by the anisotropy of the flow field.
Fig.~\ref{flow-y-fig}(c) illustrates the $x$ and $y$ components of the flow field. Although the component of the velocity along $y$ is much smaller than that along $x$, its behaviour can be related to the skyrmion configuration transition. We note that $\langle \vert u_y\vert\rangle$ increases with $\langle u_x \rangle$ below the transition indicating that part of the fluid is diverted in a direction perpendicular to the main flow, as a result of the coupling with the skyrmion’s director field. The snapshots shown in Fig.~\ref{flow-y-fig} (a) and (b) show that, in the skyrmion region, the flow lines (blue arrows) are predominantly oriented perpendicular to the director (grey lines). The magnitude of the velocity along $y$, $\langle \vert u_y\vert\rangle$ reaches a maximum at the configuration transition ($\langle u \rangle_{th}\approx45.96\, \mu$m/s) and then decreases in the ``C'' shape region.

\subsection{Two skyrmions}
\label{2sk-sec}

In the previous sections, we simulated single skyrmions on a square domain subject to periodic boundary conditions, which is equivalent to simulating a lattice of equally spaced skyrmions. The single skyrmions interact only when their size is close to the size of the simulation domain. In what follows, we report the results of simulations of a pair of moving skyrmions whose center-to-center vector is at an angle with respect to the flow direction. Before applying the external flow we allow the system of two skyrmions to relax to the configuration corresponding to the minimum Frank-Oseen elastic free energy. In the linear (far-field) approximation the  skyrmions experience an elastic repulsive interaction with dipolar symmetry. In what follows, however, we will focus on near-field effects. More specifically, we consider how the presence of a second skyrmion affects the shape of a neighbouring flowing skyrmion, compared to the shape of the single skyrmion described above.

\begin{figure}[h]
\center
\includegraphics[width=\linewidth]{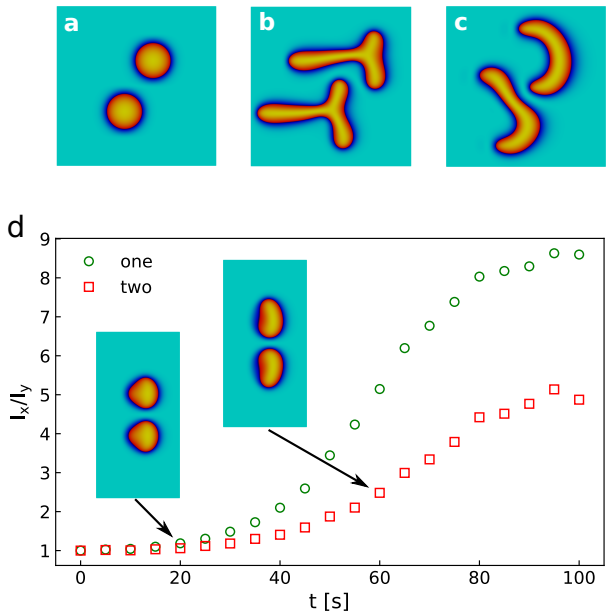}
\caption{Two flowing skyrmions. (a) Initial configuration. (b) $\langle u\rangle = 29.4 \,\mu$m/s at $t=119$ s. (c) $\langle u\rangle = 117.6\, \mu$m/s at $t=69$ s. (d) Time evolution of the order parameter for one skyrmion and two skyrmions aligned in the direction perpendicular to the flow, as shown in the insets, at two instants of time for a flow velocity $\langle u\rangle = 58.82\, \mu$m/s. }
\label{two-fig}
\end{figure}

We recall, that the shapes in the single skyrmion case exhibit mirror symmetry with respect to the $x$-axis (in the frame of reference, where the skyrmion center of mass is located on that axis). This mirror symmetry is broken (Figs.~\ref{two-fig} (b) and (c)) for two skyrmions which initially are not aligned in the direction perpendicular or parallel to the flow (Fig.~\ref{two-fig}(a)). Additionally the shapes of the two skyrmions become quite different as shown in Figs.~\ref{two-fig} (b) and (c). This shape asymmetry can be understood in terms of the flow and the inter-skyrmion repulsion arising from the elastic free energy. Indeed, as the flow induces stretching of a given skyrmion in the direction of its neighbor, the skyrmions repel, which in turn suppresses the stretching. This is noticeable when the skyrmions ``surface-to-surface'' (where the surface is defined as an iso-surface corresponding to $n_z=0$) distance becomes comparable to the cholesteric pitch.

Next, we quantify the relative change in the skyrmion shape induced by the presence of the second skyrmion. To this end, we consider a pair of skyrmions whose center-to-center vector is initially aligned with the $y$-axis (see the left inset in Fig.~\ref{two-fig}(d)), i.e., perpendicular to the direction of the external flow. We calculate the shape order parameter $I_x/I_y$, defined in Eq.~(\ref{ix-it-eq}), of one of the skyrmions of the pair and compare it to the corresponding order parameter  for a single flowing skyrmion. The order parameters are plotted in Fig.~\ref{two-fig}(d) against time. The average flow velocity is set above the configuration-transition threshold $\langle u \rangle_{th}$ for a single skyrmion, and thus both skyrmions tend to stretch in the direction perpendicular to the flow. The stretching is strongly suppressed, however, due to the elastic repulsion between the two skyrmions (see the right inset in Fig.~\ref{two-fig}(d)). For instance, at $t=20$s the surface-to-surface distance between the skyrmions is $5.9\,\mu$m and the relative difference between the order parameters is $10.6\%$ while, at $t=60$s the distance is $5.0\,\mu$m and the relative difference of the order parameters is $51.8\%$. 

An experimental setup to study the interaction between flowing skyrmions may consider a cluster of them as in Refs.~\cite{PhysRevLett.120.197203, C9SM02312G}. Depending on the distance between one skyrmion and its neighbours, the suppression of the skyrmion’s elongation should be stronger than in the two skyrmions case. A question to be addressed in future work is how the interaction between skyrmions affects the threshold velocity for the configuration transition.

\section{Conclusion}
\label{conclusion-sec}

The investigation of flowing skyrmions is an important research topic, both theoretically as well as in applications, e.g. in those involving microfluidics, as mass flows are expected to affect the skyrmions structure and stability. While this may be beneficial 
in some applications it will be detrimental in others. In any case it is necessary to identify and quantify these effects and to search for efficient ways of controlling them.   

Simulation studies of flowing skyrmions are scarce (or non-existent) and the results of the investigation reported here revealed one surprise and highlighted some of the difficulties that lie ahead. 

As for the surprise we found that for velocities above a certain threshold, 2D skyrmions stretch in the direction perpendicular 
to the flow, by contrast to the regime of weak flows where the skyrmions stretch along the streamlines of the flow field. We also found that the two regimes are separated by an abrupt (first-order) dynamical transition, which is robust with respect to e.g., changes of the liquid crystal elastic constants. 

This result clearly illustrates that simulations provide an ideal tool to make progress in the field and our study is likely to be followed by more sophisticated and realistic ones. Beyond the obvious generalizations to 3D skyrmions, a full understanding of the mechanism that drives the configuration transition is probably the most pressing question. A detailed analysis of the flow fields and of the forces acting on the system is underway and may reveal a dominant mechanism or mechanisms.

A somewhat (un)related line of research concerns the (ir)reversibility of the flowing skyrmion configurations, when the flow is reversed. Preliminary results show that, as expected, in a given flow regime the shape changes are reversible during the initial stages of the flow, and become irreversible as the flow proceeds. A quantitative analysis of the irreversibility requires, however, the calculation of the different contributions to the dissipated energy, e.g. dissipation from the fluid flow, dissipation from the director relaxation, dissipation from the anchoring conditions, which will be carried out in future work, both in the different flow regimes as well as near the skyrmion configuration transition.This may shed further light on the dominant mechanism of the configuration transition and provide a means to control skyrmion shapes.

\section*{Appendix: Parameters}

Table~\ref{tab1} summarises the parameters used in the simulations in numerical and physical units. The conversion between the two systems is obtained from the values of $\Delta t$, $\Delta x$ and $\rho$. We note that the value of the time step is only a reference, since we used a time step $100 \Delta t$ to solve Eq.~(\ref{director-time-eq}) with finite-differences. In addition, we recall that we used elastic constants smaller than those of MBBA.
\begin{table}
\caption{\label{tab1} Parameters used in simulation and physical units.}
\footnotesize
\begin{tabular}{@{}llll}
\br
symbol&sim. units & physical units&description\\
\mr
$\rho$&1&1088 Kg/m$^{3}$&density\\
$\Delta x$&1 & 0.3125 $\mu$m& lattice spacing\\
$\Delta t$&1 & 10$^{-6}$ s& time step\\
$K_{11}$&0.01&9.54$\times 10^{-14}$ N &elastic constant\\
$\alpha_4$&819.2&0.08 Pa.s&Leslie viscosity\\
$c$ & 10 & 0.3125 m/s & see Eq.~(\ref{pseudo-time-eq2})\\
$p$ & 32 & 10 $\mu$m & cholesteric pitch\\
$W_0$ & 0.00012 & 3$\times 10 ^{-5}$ J/m$^2$& anchoring strength\\
$\chi$ & 4.25 & 3$\times 10^{-11}$ N s/m & friction coefficient\\
\br
\end{tabular}\\
\end{table}
\normalsize

\section*{Acknowledgments}
We acknowledge financial support from the Portuguese Foundation for Science and Technology (FCT) under the contracts: IF/00322/2015, PTDC/FIS-MAC/5689/2020, UIDB/00618/2020 and UIDP/00618/2020.

We are also thankful to  Ivan Smalyukh for the suggestions and fruitful discussions.
\\\\

\bibliography{refs} 
\bibliographystyle{unsrt} 

\end{document}